\documentstyle[11pt,paspconf,psfig]{article}

\def\etal{et al.}
\def\hii{H{\sc ii}}

\def\msun{M$_{\odot}$}
\def\zsun{Z$_{\odot}$}
\def\halpha{\ifmmode {\rm H{\alpha}} \else $\rm H{\alpha}$\fi}
\def\hbeta{\ifmmode {\rm H{\beta}} \else $\rm H{\beta}$\fi}
\def\heii{\ifmmode {{\rm He{\sc ii}} \lambda 4686} \else {He\,{\sc ii} $\lambda$4686}\fi}
\def\Heii{He\,{\sc ii}}

\def\niii{N\,{\sc iii} $\lambda$4640}

\def\ciii{C\,{\sc iii} $\lambda$4650}

\def\ciiib{C\,{\sc iii} $\lambda$5696}
\def\civ{C\,{\sc iv} $\lambda$5808}

\begin{document}

\title{Models for WR and O Star Populations in Starbursts: 
New Developments and Applications to Large Samples of WR Galaxies} 

\author{Daniel Schaerer} 
\affil{Observatoire Midi-Pyr\'en\'ees, 
       14, Av. E. Belin, F-31400, Toulouse, France
      (schaerer@obs-mip.fr)}


\begin{abstract}
We summarise recent developments on synthesis models for massive star 
populations with a particular emphasis on Wolf-Rayet (WR) stars. 
Quantitative analysis of the stellar content of WR galaxies are reviewed.
Comparing observations of WR galaxies from various samples with synthesis
models we derive constraints on their burst properties. 
The observations indicate very short burst periods and are generally 
compatible with a Salpeter IMF and a large upper mass cut-off. 
The use of the \hbeta\ equivalent width as an age indicator works well
for WR galaxies.
We briefly summarise comparisons of stellar populations in super star clusters 
which also provide useful contraints on evolutionary models for massive stars 
e.g.\ at very low metallicities inaccessible in the Local Group.
In particular the observed WN and WC populations favour the high mass loss models
of Meynet et al. (1994).
Finally we review recent work on the origin of nebular \Heii\ emission.
From the new catalogue of WR galaxies and high excitation \hii\ regions of
Schaerer et al. (1998) we find a close relation between the presence of
nebular \Heii\ and WR stars. 
The analysis of individual WR galaxies including I Zw 18 supports the suggestion
of Schaerer (1996) that hot WR stars are responsible for the hard ionizing
flux.

\end{abstract}


\section{Introduction}

\section{Synthesis models for young starbursts}
There exist several independent synthesis models suited to the
analysis of massive star populations in WR galaxies in particular
(cf.\ review of Leitherer, these proceedings).
A subset of models predict essentially the relative distributions
of different populations (e.g.\ WN, WC, O stars), SN rates etc.\
(e.g.\ Meynet 1995, Vanbeveren \etal\ 1997, 1998), whereas no (or very
limited) direct predictions about the spectral properties or other observables
are made. The models of Cervi\~no \& Mas-Hesse (1994), Leitherer
\& Heckman (1995) or the updated/extended version ({\em Starburst99:}
Leitherer \etal\ 1999), Schaerer (1996) and Schaerer \& Vacca (1998),
on the other hand predict numerous observational features including
at least the hydrogen recombination lines and the strength of WR feature(s),
essential to the interpretation of the massive star content in young
starbursts.

For the following we shall rely on the recent synthesis models
of Schaerer \& Vacca (1998, hereafter SV98) who predict predict 
the strongest WR lines in the UV and optical:
\Heii\ $\lambda$1640, \niii, \ciii, \heii, \Heii$+$\hbeta\
$\lambda$4861, \Heii\ $\lambda$5412, \ciiib, \civ, \Heii$+$\halpha\
$\lambda$6560.
The contribution from WNL, WNE, WC4-9, and WO stars to these lines
is taken into account in the models based on newly determined empirical 
line luminosities (see SV98). The \heii\ emission from Of stars
and Of/WN stars found in young clusters (e.g.\ R136, NGC 3603, cf.\ 
Heap, Drissen, these proceedings) is also considered 
(see SV98, Schaerer \etal\ 1999, hereafter SCK99).
The SV98 models allow in particular direct comparisons with the \heii\
line which is preferred over the use of the so-called broad ``WR bump'', 
well known to consist in general of a blend of several stellar (WR) 
but also nebular lines rendering the interpretation of its strength 
difficult.
The use of other independent WR lines (e.g.\ \ciii, \civ) now observed 
increasingly often (cf.\ SCK99, compilation of Schaerer \etal\ 1998)
furthermore allows to distinguish between WN and WC subtypes, 

An important ``feature'' of these synthesis models is that they rely
on stellar evolution tracks (Geneva models) which have extensively been 
compared with observations of both individual WR stars in the 
Galaxy and the LMC as well as WR and O star populations in the Local Group
(Maeder \& Meynet 1994). These comparisons in particular favour the
high mass loss models (Meynet \etal\ 1994) generally used here. 
It is understood that although the adopted mass loss rates in the WR 
phases are likely too high compared
to observations including the effects of clumping (e.g.\ Hillier, these
proceedings) this effect may be compensated by additional mixing
processes leading to a similar prolongation of the WR phase (cf.\ 
Maeder, Meynet, these proceedings).

\section{Analysis of WR and O stars in starbursts}
The latest catalogue of WR galaxies (Schaerer \etal\ 1998, hereafter SCP98) 
includes now 
nearly 140 objects (mostly giant \hii\ regions, BCD, etc.\ but also 
some starburst nuclei, IRAS galaxies etc.)
However, so far there have been fairly few {\em quantitative} analysis 
of the WR and O star content and the properties of these regions.
The aim of this Section is to summarise previous work and extend 
currently existing studies to a larger sample.

\begin{figure}[ht]
\centerline{\psfig{figure=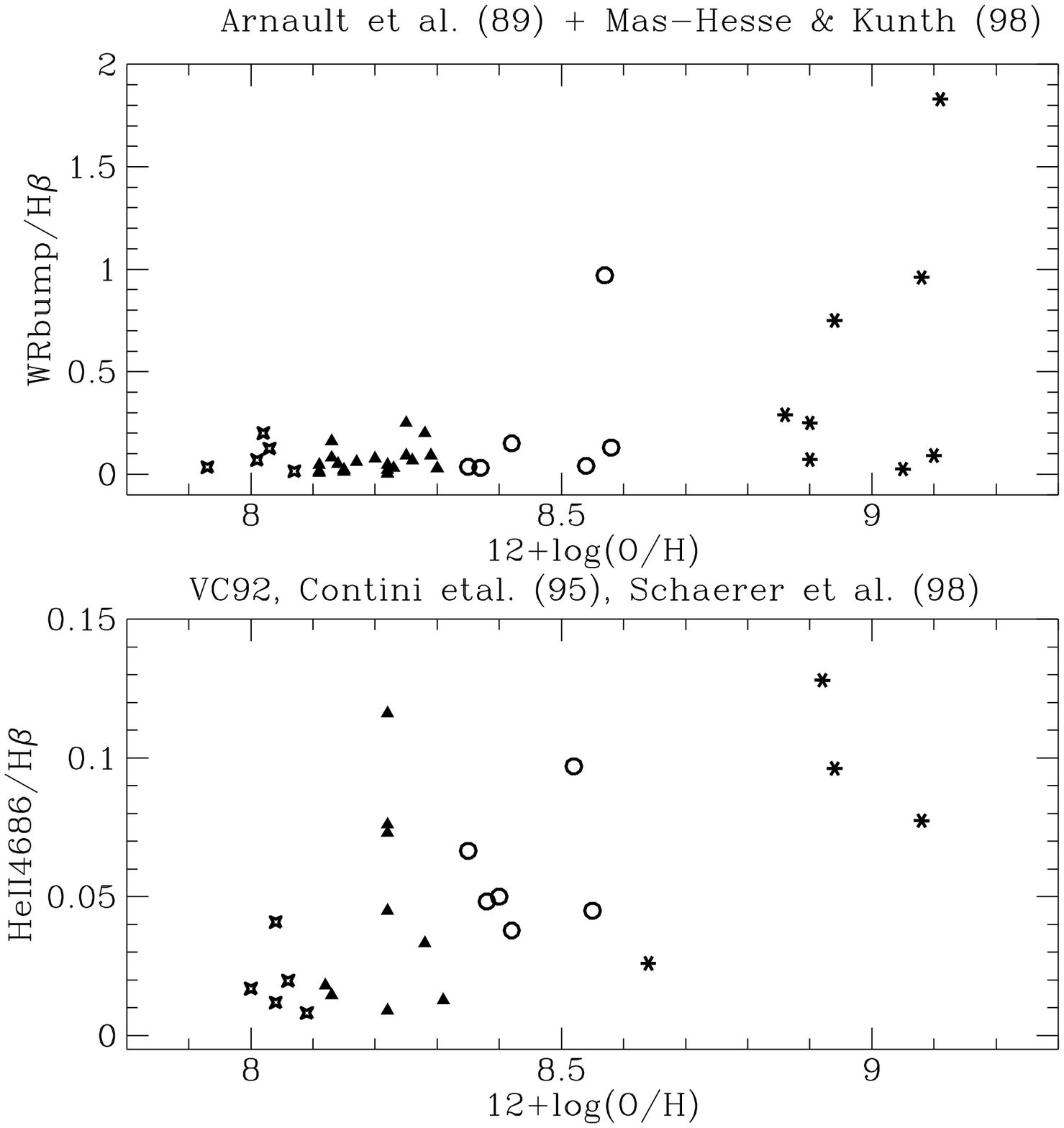,width=7cm}
\psfig{figure=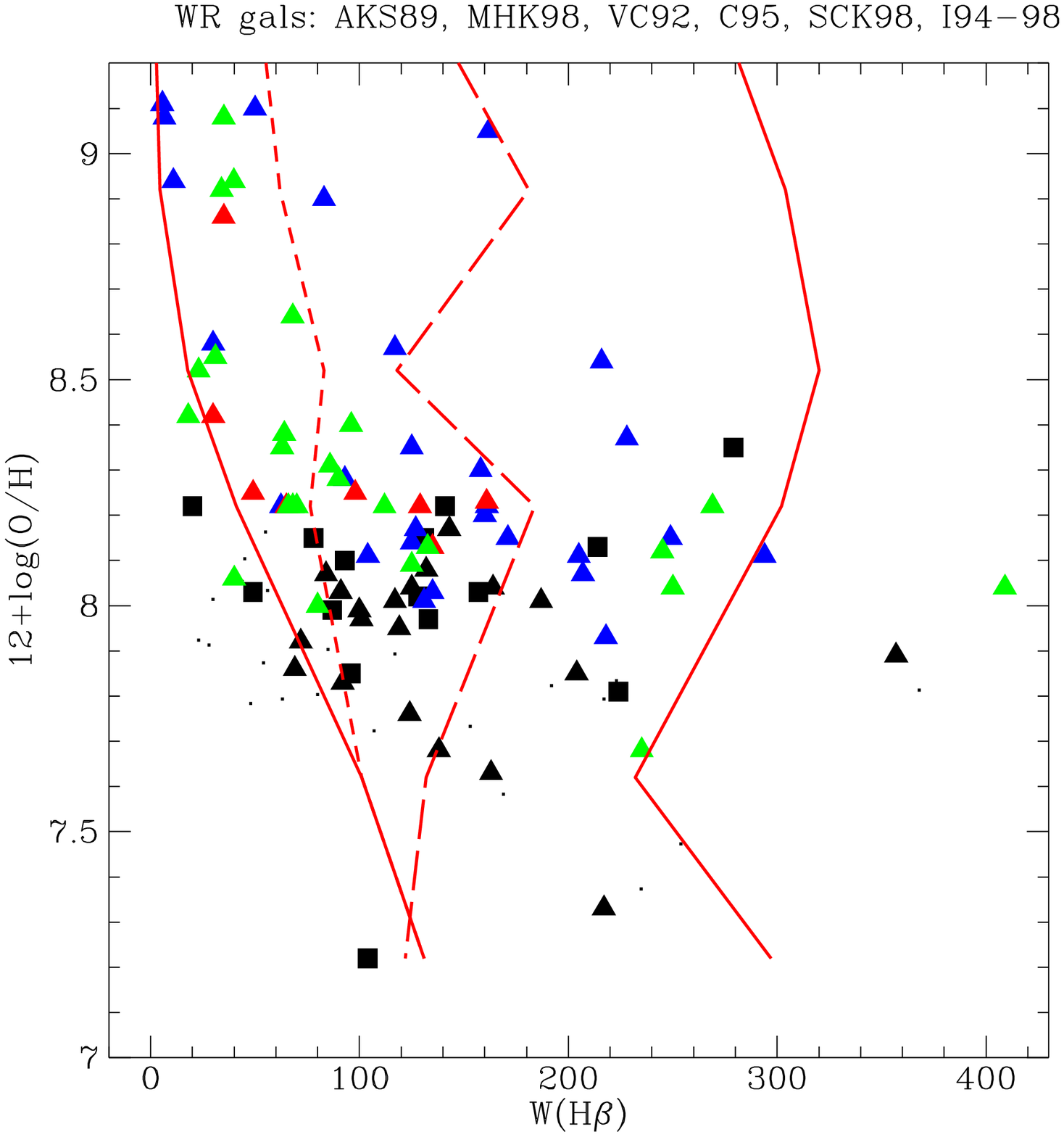,width=7cm}}
\caption{{\bf Left --} {\em Top:} WRbump/\hbeta\ intensity as a fct.\ of
O/H for the objects from Arnault \etal\ (1989) and Mas-Hesse \& Kunth (1998).
{\em Bottom:} Same for \heii/\hbeta\ from the samples of Vacca \& Conti (1992),
Contini \etal\ (1995) and SCK99.
Different symbols represent different metallicity ``bins'' (see Fig.\ 
\protect\ref{fig_heii_whb}).
{\bf Right --} Position of WR galaxies in metallicity (O/H) vs.\ \hbeta\ equivalent
width (triangles, squares). Different greytones/colours indicate different samples.
Solid lines denote the beginning/end of the WR-rich phase predicted by the
SV98 models. Dashed lines indicate age steps of 1 Myr during the WR phase.
}
\label{fig_combine}
\end{figure}

The bulk of data available in the literature on the WR signature in 
integrated spectra is summarised in Fig.\ \ref{fig_combine} (left), showing
the relative intensity of the WR bump (\heii) in the top (bottom) panel
as a function of metallicity.
In the early study of Arnault \etal\ (1989) the top figure was already
interpreted as showing the following:
{\em 1)} The spread in WR/\hbeta\ at a given O/H is due to different 
  ages.
{\em 2)} There is an increase of the envelope of WR/\hbeta\ with O/H,
which can be simply understood by the decrease of the mass limit for 
WR formation towards higher metallicity already shown by Maeder \etal\
(1980). 
{\em 3)} Their first quantitative modeling also indicated that {\em short}
star formation periods seemed to be required to achieve the large observed 
WRbump intensities.

Several subsequent studies have secured spectra with sufficient resolution
and S/N to allow a reliable measurement of the broad \heii\ line 
in particular
to circumvent the above mentioned difficulties associated with the WR bump.
In passing we note from Fig.\ \ref{fig_combine} (left, bottom) that \heii/\hbeta\ 
from these samples shows the same trends as discussed above.
Kunth \& Sargent (1981),
VC92 and Vacca (1994) put forward the determination
of the number of WN and equivalent O stars as a convenient way to 
quantify the massive star populations. Although affected by several uncertainties
and systematic effects (cf.\ below) this simple method is still often used.
Comparisons of the WN/O number ratios of VC92 with simple ``number'' synthesis 
models clearly show that very short
($\sim$ instantaneous compared to the lifetime of massive stars) bursts of 
star formation are required to reproduce the large WN/O values
(e.g.\ Maeder \& Conti 1994, Meynet 1995). The observed variation of WN/O
with metallicity is also fairly well reproduced by these models.

The major drawback of quantitative comparisons of ``observed'' WN/O 
number ratios with evolutionary models stems from the determination
of the number of O stars (from a H recombination line), 
which is dependent on the IMF, star formation history, and age of the
population (cf.\ Vacca 1994, SV98).
Indeed as shown by Schaerer (1996) the method of VC92 and Vacca (1994) 
leads to a systematic {\em underestimate} of the number of O stars.
Comparisons of WR/O number ratios must therefore be taken with caution
e.g.\ when used to constrain the IMF slope 
(cf.\ Contini \etal\ 1995, Schaerer 1996).

It is encouraging that the disadvantage of the previous method can be 
easily circumvented without the need to invoke any additional major
assumption. Evolutionary synthesis models (such as those discussed
above) can directly predict the observable
quantities (\heii\ and \hbeta\ line intensities) for given SFR, IMF and age 
-- the assumption relating the ionizing photon flux to \hbeta\ (e.g.\ case 
B recombination) remains as before.
Comparisons with observations can thus directly be made in a plane
of observables. Results from such studies are presented in the
next section.

\subsection{Comparisons with recent synthesis models}

{\bf Uncertainties:} Before we proceed it is useful to remember
the most important potential observational and theoretical uncertainties
involved in this process (cf.\ Conti, these proceedings).
The measurement of relative line WR/\hbeta\ line intensities
and equivalent widths can mostly be affected by 3 effects (see SCK99):
{\em (i)} Not all ionizing photons are counted (e.g.\ small slit,
leakage etc.).
{\em (ii)} Stars and gas suffer from a different extinction.
{\em (iii)} An underlying older population contributes additional 
continuum light.
Some uncertainties due to the synthesis models (stellar tracks excluded)
are (SCK99):
{\em (a)} The use of the intrinsic WR line luminosity which shows
e.g.\ a relatively large scatter for \heii\ (SV98, Fig.\ 1).
{\em (b)} Interpolation of tracks leading to uncertainties in the relative
WC to WN populations.
{\em (c)} Assumptions on the calculation of H recombination lines.

{\bf On the W(\hbeta) age indicator for WR galaxies:} 
In Fig.\ \ref{fig_combine} (right panel) we have plotted the position
of all WR galaxies from the samples of Arnault \etal, Mas-Hesse \& Kunth
(1998), VC92, Contini \etal\ (1995), SCK99, and
Izotov and collaborators (see references in Izotov \& Thuan 1998)
in the W(\hbeta)-metallicity plane. 
Note that the bulk of these objects represent star forming regions
(e.g.\ giant \hii\ regions etc.) in BCDs or spirals.
If the \hbeta\ equivalent width is taken as an age indicator, the 
temporal evolution proceeds to the left in this plot.
Overplotted are the predictions from the SV98 models
for W(\hbeta) at the beginning and end of the WR phase for instantaneous
burst models with a Salpeter IMF and $M_{\rm up}=$ 120 \msun.
We note that the essentially all WR galaxies lie within the predicted
W(\hbeta) range; in the most populated metallicity range (12+$\log$(O/H)
$\sim$ 7.8 -- 8.5) the observations also fully populate this domain.
Although it must be noted that each individual observation of W(\hbeta) may 
be subject to the uncertainties mentioned above, we conclude
from Fig.\ \ref{fig_combine} (right) that on average the 
{\em age and duration of the WR-rich phase predicted by the SV98 models 
for instantaneous bursts agree quite well with the observations.}
In this context it must, however, be reminded
that regions with very 
large \hbeta\ equivalent widths such as predicted for populations of 
ages $\sim$ 0-1 Myr are not observed. The origin of this well known 
discrepancy is not understood yet.


\begin{figure}[htb]
\centerline{\psfig{figure=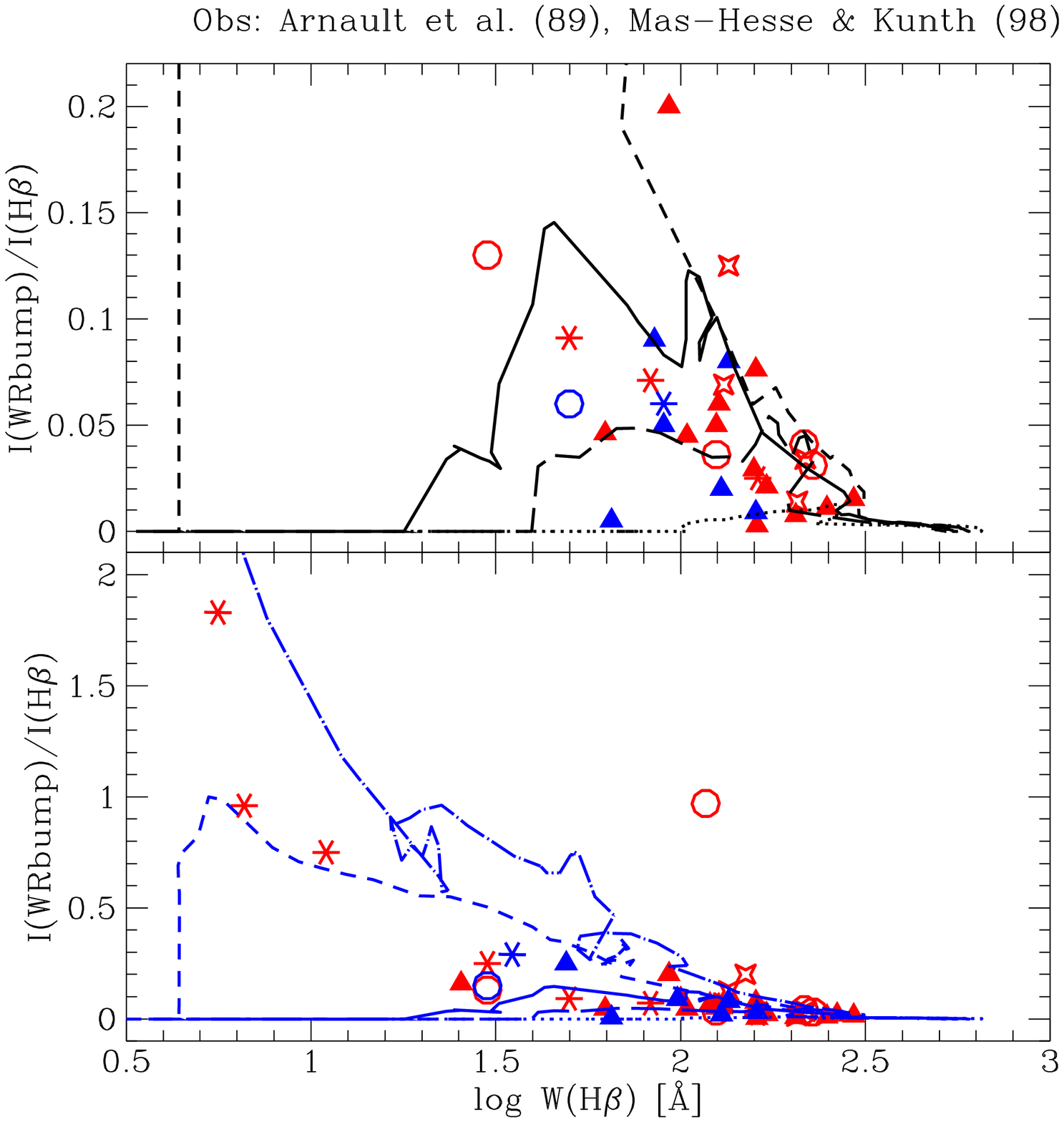,width=7cm}
\psfig{figure=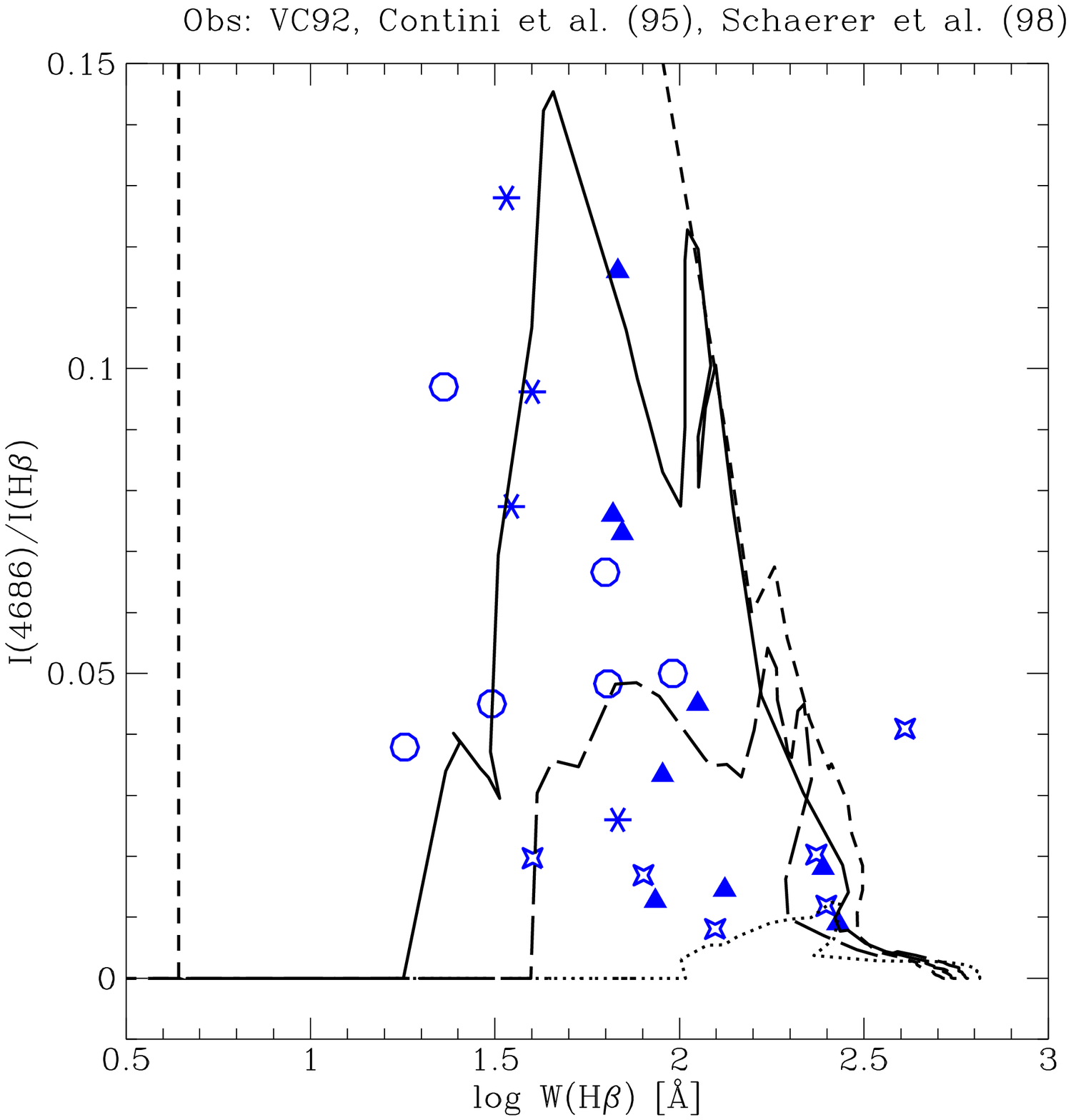,width=7cm}}
\caption{{\bf Left --} Comparison of relative WRbump/\hbeta\ intensities 
from various samples (see text) with synthesis models of SV98 
(instantaneous burst, Salpeter IMF). The upper panel shows a blow-up of 
the bottom.
{\bf Right --} Same as left for \heii/\hbeta. Samples described in
Sect.\ 3.2.
In all Figs.\ metallicities are indicated by different symbols:
Stars ($Z >$ 0.01) to be compared with dashed and dott-dashed lines 
(models at $Z=$0.02 and 0.04 resp.),
circles (0.005 $< Z \le$ 0.01) to be compared with solid line ($Z=$0.008 models),
filled triangles (0.003 $< Z \le$ 0.05) vs.\ long-dashed ($Z=$0.004 models),
4-star ($Z <$ 0.003) vs.\ dotted (($Z=$0.001 models).
}
\label{fig_heii_whb}
\end{figure}

{\bf Observations of the WR bump:} 
In Fig.\ \ref{fig_heii_whb} (left) we have plotted the WRbump/\hbeta\
intensities from the samples of Arnault \etal\ (1989) and Mas-Hesse
\& Kunth (1998). When two measurements are available we used
more recent observations of Mas-Hesse \& Kunth; also the WRbump 
intensities corrected for the underlying population and differential
extinction were used (see Mas-Hesse \& Kunth 1998).
Note also that higher resolution spectra measuring \heii\ are available
for many of these objects (see below).
Comparing these observations with the model predictions 
at the appropriate metallicity shows the following:

{\bf\em 1)} Most objects have WRbump intensities around or below the model 
prediction indicating that the observed WR and O star populations are reasonably
well reproduced by the models calculated for instantaneous bursts.

{\bf\em 2)} 
The largest WRbump/\hbeta\ are found in objects with higher metallicity.
This can be understood in the following way:
The strong WR/\hbeta\ increase is mostly due to the decrease of \hbeta;
the equivalent width of the WR features increase much more slowly (cf.\ SV98).
The rapid decrease of \hbeta\ in turn is naturally predicted by
evolutionary models at solar and higher metallicity, due to a cooler
ZAMS and an extension of the MS to lower temperatures (see SV98).
Such a metallicity effect on MS stellar evolution is also supported
by studies of the nebular properties of \hii\ galaxies
(e.g.\ Garc\'{\i}a-Vargas \etal\ 1995, Stasi\'nska \& Leitherer 1996).
Other effects like dust could, however, also provoke such changes with 
metallicity as pointed out by T.\ Heckman during this meeting.
Studies of larger samples of high metallicity regions and detailed
multi-wavelength analysis would be of interest.

{\bf\em 3)} A few objects show WRbump intensities largely in excess of the
model predictions\footnote{E.g.\ in Fig.\ \ref{fig_heii_whb}
 NGC 1741: ($\log$ W(\hbeta),WR/\hbeta) $\sim$ (2.,0.2);
Tol 89: (2.1, 0.13); Mrk 4038 \#3: (2., 1.)).}.
For some of them (e.g.\ NGC 1741) later measurements of \heii\ (see below)
show no discrepancy. Other data (e.g.\ from Rosa \& D'Odorico 1986) is quite 
uncertain. There remain, however, some objects with high quality measurements
and an ``excessively'' strong \heii, which might be indicative of an unusual
IMF (e.g.\ Tol 89, see SCK99 and below, also Huang \etal\ 1998).


\subsection{IMF and burst duration from individual WR lines}
In Fig.\ \ref{fig_heii_whb} (right) we plot the majority of the observations 
where a measurement of \heii\ is available.
Using this line instead of the WR bump together with additional WR lines 
(see SCK99) should allow more accurate comparisons
and provide e.g.\ even a reasonable handle on the IMF slope.
A comparison of the observations of VC92 with the SV98 models has been
presented by Schaerer (1996). New high S/N spectra of NGC 5253, NGC 3049, 
NGC 3125, He 2-10 and Tol 89 were analysed in detail by Schaerer 
\etal\ (1997) and SCK99.  Here we briefly summarise their main
results (see also Contini \etal, these proceedings).

Inspection of Fig.\ \ref{fig_heii_whb} shows that the vast majority of
the observed \heii\ intensities are well reproduced by the models at the
appropriate metallicity assuming an instantaneous burst and a Salpeter IMF.
However, some objects show large \heii/\hbeta. {\em Is this convincing evidence
calling e.g.\ for a flatter IMF ?} We think the answer is no.
As pointed out by SCK99 it is useful to compare also the equivalent widths
of the WR feature(s) since these are affected differently by the uncertainties
listed above. Indeed in some objects there is evidence showing
a different spatial distribution of the gaz and stars (SCK99): in this case 
the use of WR/\hbeta\ line intensities may not be reliable and it is preferable
to use quantities related only to the stellar content (i.e.\ W(WR)).
Furthermore the detection of several ``independent'' WR lines (\heii, \civ)
increases the number of constraints on the models.

The detailed comparisons of SCK99 taking these effects into account
yield the following:
{\em 1)} Essentially instantaneous bursts ($\Delta t \la$ 2-4 Myr) are required 
to reproduce the observed WR line intensities. 
{\em 2)} The majority of the observations can be reproduced with a Salpeter IMF.
No clear case requiring a flatter IMF is found. 
{\em 3)} The IMF in these regions must be populated up to large masses.
For the case of I Zw 18, at $Z \sim$ 1/50 \zsun, $M_{\rm up}$ extending
to 120-150 \msun\ is found (de Mello \etal\ 1998).
{\em 4)} The relative populations of WN and WC stars detected clearly favour
the high mass evolutionary models (SCK99).

We note that our results on the IMF slope and the upper mass cut-off are in 
agreement with several independent studies, e.g.\ stellar counts
(review by Massey 1998), UV line profile modeling of starbursts
(references in Leitherer 1998), photoionization models for
\hii\ galaxies (e.g.\ Garc\'{\i}a-Vargas \etal\ 1995, Stasi\'nska \& 
Leitherer 1996) etc.
Interestingly, however, there are several studies indicating the possibility
of a lower value for $M_{\rm up}$ at metallicities above solar
(e.g.\ ULIRG: see review of Leitherer 1998, HII regions: Bresolin \etal\ 1998).
A search for WR stars in metal rich regions should allow to put direct 
constraints on the upper end of the IMF in such environments.

\section{The origin of nebular HeII emission}
The presence of nebular \heii\ emission indicative of high excitation
in extra-galactic \hii\ regions and WR galaxies has attracted new attention
in recent years. Here we shall briefly summarise the current status 
and recent work aiming at understanding the nature of this emission
(see also Schaerer 1996, 1998).
These studies have a direct bearing on our understanding of the properties
of WR stars, the structure and physics of \hii\ regions, and our knowledge
of the ionizing fluxes of starbursts. Some of these aspects are also
discussed by Crowther, Galarza \etal, Pen\~a \etal, Stasi\'nska and others
in these proceedings.

{\bf Empirical facts:}
In the Local Group eight nebulae exhibit nebular \heii\ emission indicative of 
unusually high excitation. All except one Galactic object are located in 
low metallicity environments (IC 1613, SMC, LMC). 
Except for two cases the nebulae are associated with early WN and WO stars.
Approximately 60 objects (mostly giant \hii\ regions) outside the Local Group
with nebular \heii\ are now known (see compilation of Schaerer \etal\ 1998, SCP98).
Most of them have low metallicities, and typically I(4686)/I(\hbeta) $\sim$
1 -- 5 \%.
Many objects where originally {\em nebular} \heii\ was reported, have revealed the
presence of {\em broad} 4686 in spectra with better S/N and resolution, and are
therefore classified as WR galaxies. Many WR galaxies also exhibit simultaneously
nebular \heii\ emission.
These observations strongly suggest that the phenomenon of nebular \Heii\ emission
is related to the presence of WR stars.

{\bf Quantitative models:}
In this context Schaerer (1996) presented the first models aiming at a quantitative
explanation of nebular \Heii\ emission in WR galaxies and related objects.
Indeed it is found that synthesis models using appropriate model atmospheres
for WR and O stars (see also SV98) are able to produce nebular \Heii\ with 
intensities in the range of the observations during a relatively short phase 
where hot WN and WC/WO stars are present. This is in agreement with the presence
of nebular \heii\ in several WR galaxies (see Schaerer 1996, Mas-Hesse \& Kunth
1998).
It is also supported by the detection of WR stars in I Zw 18
(Legrand \etal\ 1997, Izotov \etal\ 1997), the galaxy with the lowest metal 
content known: using appropriate stellar tracks one is able to reasonably
reproduce the WR content and nebular \heii\ emission in this galaxy
(de Mello \etal\ 1998, Stasi\'nska these proceedings).
At intermediate to high metallicities \heii\ is predicted to be largely dominated
by stellar emission from WR stars (Schaerer 1996, SV98).
In addition, compared to the atmosphere models used in SV98 the blanketing effects 
discussed by Crowther (this meeting) are also likely to progressively reduce the 
high energy photon flux of WR stars with increasing metallicity.
Under reasonable assumptions and using evolutionary synthesis models one can also 
rule out several other potential sources of \heii\ emission
(cf.\ Garnett \etal\ 1991): Of stars, O stars close to the Eddington limit,
and X-ray binaries.
We conclude that synthesis models including appropriate model atmospheres
are able to reproduce the observed nebular \heii\ emission when WR stars
are present.

Independent information supporting WR stars as the origin of \Heii\ ionizing photons
comes from (admittedly scarse) spatial information:
When observed the spatial distribution of nebular \heii\ follows that of the 
WR features (Legrand \etal, Izotov \etal, Maiz-Apellaniz \etal, de Mello \etal). 
Statistically one can also make the following arguments (see samples in SCP98):
{\em 1)} In the sample of Izotov and collaborators 40 objects show nebular \Heii;
 75 \% of them show also WR features (half of them WN and WC). Conversely, from 
the 18 objects {\em without} nebular \Heii\ only 28 \% harbour WR stars.
{\em 2)} In the total sample of SCP98 aiming at a complete list of \heii\ detections
from the literature, 54 objects exhibit both broad (WR) and nebular 4686, whereas 
only 19 \hii\ regions show purely nebular \Heii.
Clearly the vast majority of objects showing nebular \Heii\ harbour WR stars.
This corroborates the results discussed above.

Analysis of the Local Group objects allowing a detailed investigation
of the stellar content and the physical processes in the nebula would be 
rewarding to test these results. Further studies of the \hii\ regions with purely 
nebular \Heii\ need also to be undertaken.

\section{Conclusions}
We have summarised recent developments on synthesis models for massive star 
populations with a particular emphasis on WR stars. The SV98 models rely in particular
on up-to-date atmospheres and stellar tracks which have extensively been
compared to observations in the Galaxy and the Local Group.
Comparisons of observations of WR galaxies with these and other synthesis models
yield the following main results:
{\em 1)} The predicted age and duration of the WR-rich phase agrees well 
with the observations. 
{\em 2)} Observations of high WRbump/\hbeta\ intensities in high metallicity objects
seem to be due to a low \hbeta\ flux. This can be explained by metallicity effects
on the stellar tracks although other explanations are also possible.
{\em 3)} The observed WR and O star populations in all objects indicate nearly
instantaneous bursts. 
{\em 4)} The vast majority of the objects can be reproduced with
a Salpeter slope for the IMF. The upper mass cut-off is large in these galaxies
such as to produce enough WR stars.
{\em 5)} The quantitative analysis of both WN and WC populations in WR galaxies
supports high mass loss evolutionary tracks. Other mechanisms (e.g.\ rotational
mixing) may have similar effects.
{\em 6)} As shown by the example of I Zw 18 (de Mello \etal\ 1998), 
the analysis of stellar populations 
in super star clusters may provide useful constraints on the evolution of massive 
stars, especially at low metallicities inaccessible in the Local Group.

The origin of nebular \Heii\ emission requiring copious ionizing photons with
energies above 54 eV has been puzzling.
We find that the observations in the Local Group and more distant objects 
suggest a close relation between the presence of nebular \heii\ emission and WR stars.
Quantitatively the observed nebular \Heii\ can be explained 
by synthesis (and photoionization) models due to the presence of hot WN and/or WC/WO
stars (Schaerer 1996, de Mello \etal, Stasi\'nska these proceedings).

The work summarized here should improve our knowledge not only on the evolution and 
atmospheres of massive stars, but also on starbursts and their properties.

\begin{acknowledgements}
DS acknowledges a grant from the Swiss National Foundation of Scientific
Research and financial support from the IAU and the GdR Galaxies.
\end{acknowledgements}

\end{document}